# A Path to Smart Radio Environments: An Industrial Viewpoint on Reconfigurable Intelligent Surfaces

Ruiqi Liu, *Member, IEEE*, Qingqing Wu, *Member, IEEE*, Marco Di Renzo, *Fellow, IEEE*, and Yifei Yuan, *Senior Member, IEEE*

*Abstract*—With the standardization and commercialization completed in an unforeseen pace for the 5th generation (5G) of wireless networks, researchers, engineers and executives from academia and industry have turned their sights to new candidate technologies that can support the next generation of wireless networks. Reconfigurable intelligent surfaces (RIS), sometimes referred to as intelligent reflecting surfaces (IRS), have been identified as a potential component of future wireless networks, because they can reconfigure the propagation environment by using low-cost nearly-passive devices. With the aid of RISs, the coverage of a cell can be expected to significantly increase, and with it the overall throughput of the network. RISs have not only become an attractive research area but have also triggered several collaborative research projects that aim to develop innovative algorithmic solutions and hardware demonstrators and prototypes. In parallel, technical discussions and activities towards standardization have already taken off in some countries. Promoting RISs to be integrated into future commercial networks and to become a commercial success require significant standardization efforts that need to take place at regional-level standards developing organizations (SDO) and international-level SDOs, such as the 3rd Generation Partnership Project (3GPP). While many research papers study how RISs are optimized and what their ultimate performance limits are, little effort has so far been devoted to analyze the challenges to commercialize RISs and how RISs can be standardized. This paper intends to shed some light on RISs from an industrial viewpoint and to provide a roadmap in order to make RISs industrially feasible.

## I. INTRODUCTION

Although fifth-generation (5G) wireless networks have been gradually deployed worldwide, the key enabling 5G technologies face several challenges for their practical implementation. For example, telecommunication operators in China have recently announced to shut down some 5G base stations (BSs) equipped with massive multiple-input multiple-output (MIMO) antennas [1]. This is because massive MIMO requires the use of a large number of active transmit radio-frequency (RF) chains, leading to high hardware cost and significant energy consumption. Furthermore, shrinking the cell size for supporting the increasing data traffic needs more and more active BSs and access points (APs), so as to enhance the network coverage, which further aggravates the expenditure cost of telecommunication operators for maintenance and gives rise to other practical issues such as backhaul and interference management. On the other hand, exploiting higher frequency bands, such as the millimeter wave (mmWave) and terahertz (THz) frequencies, with a much shorter wavelength inevitably results in deploying even more active network components and antennas in order to compensate for the severe propagation loss over the air, which also leads to an increased signal processing complexity. Furthermore, the most recent applications, such as immersive virtual reality, connected autonomous systems, the industrial Internet of Things, pose several stringent communication requirements in terms of high data rate, reliability, latency, coverage, security, etc, which are expected to be supported by 5G advanced wireless networks. Therefore, the academia and industry are under the pressure to develop new and cost-effective technologies.

Recently, reconfigurable intelligent surfaces (RIS), sometimes also known as intelligent reflecting surfaces (IRS), has emerged as a promising new technology to address the above mentioned issues [2], [3]. Specifically, an RIS is a planar surface, typically composed of a large number of low-cost nearly-passive reflecting elements, that has the capability to shape the amplitude and phase shift of the impinging signals. This is usually achieved by using an associated smart controller, e.g., implemented with an FPGA, which communicates with the active BSs and users via a separate wireless link for coordinating transmission and exchanging information on, e.g., channel knowledge and real-time control. By smartly coordinating the scattering of all the RIS elements, a dense deployment of RISs is able to reconfigure the end-to-end wireless channels and eventually realize the vision of smart and programmable wireless propagation environments. From an implementation perspective, RISs also exhibit appealing and competitive advantages in practice. First, since RISs do not need active RF chains and only need nearly-passive scattering elements, they have orders-of-magnitude lower hardware cost and energy consumption as compared to 5G technologies based on active antenna arrays. Second, the passive scattering of RISs mimic a full-duplex (FD) transmission mode that is free of antenna noise amplification and self-interference, which makes it more attractive than traditional active half-duplex (HD) relays that suffer from low spectral efficiency or FD relays that require sophisticated techniques to cancel the self-interference. Since RISs are nearly-passive, they can be fabricated with light weight, low profile, and even made to be conformal to various objects. As a result, they can be easily deployed in a wide range of scenarios such as walls, ceilings, billboards, lampposts, and even on the surface of

R. Liu is with ZTE Corporation, Beijing, and State Key Laboratory of Mobile Network and Mobile Multimedia Technology, Shenzhen, China. Q. Wu is with the State key laboratory of Internet of Things for Smart City, Macau, China. M. Di Renzo is with Université Paris-Saclay, CNRS, CentraleSupélec, Laboratoire des Signaux et Systèmes, 3 Rue Joliot-Curie, 91192 Gif-sur-Yvette, France. Y. Yuan is with China Mobile Research Institute, Beijing, China.

vehicles to support several applications for smart factories, stadiums, shopping centers, airports, etc. Last but not the least, RISs can be deployed as energy-efficient auxiliary devices that are transparent to the wireless users, without the need of modifying the hardware configuration of the end-user devices. This offers high flexibility and compatibility with legacy wireless systems.

In this paper, we aim to present the latest progress on research, development and standardization on RISs, as well as to provide some insights on how RISs can be integrated into global standards and commercial networks. Critical challenges down this path are analyzed and possible solutions are suggested. The organization of this paper is as follows. Section II gives the latest progress in the industry and funded projects. Section III provides key use cases for RISs. The major challenges from the industrial point of view are described in Section IV. In Section V, we illustrate the current path towards the standardization of RISs. Finally, Section VI concludes the paper.

## II. Industrial progress, Projects and Standardization

RISs are a new and emerging paradigm in wireless communications but the realization of metasurface-based RISs is still in its early stage. Nevertheless, there are some proof-of-concept platforms and hardware testbeds that have been built by different teams around the world, which can substantiate and verify theories and help researchers to study RIS-aided communications. In this section, the industrial progress on RISs is introduced from four angles: prototypes, published white papers, projects supported by funding agencies, and ongoing standardization activities.

The industrial progress of testing metamaterial-based reflective surfaces mainly originated from NTT DOCOMO, a Japanese network operator. In collaboration with Metawave Corporation, NTT DOCOMO demonstrated a 5G mobile communication system in the 28 GHz band using a metasurface reflect-array in November 2018, which achieved a downlink data rate of 560 megabits per second (Mbps) in lieu of 60 Mbps when no reflector is deployed. In January 2020, NTT DOCOMO conducted a trial of a transparent dynamic metasurface with 5G radio signals in the 28 GHz band. The designed metasurface can reflect the incident radio signals with full or partial power or can let the signal penetrate with close-to-zero power losses. This trial was re-conducted in January 2021 by NTT DOCOMO with another partner, AGC Inc., by confirming the previous finding that a transparent metasurface can improve the power level of the radio signals received at indoor focal points.

Researchers from universities have contributed significantly to building and testing proof-of-concepts and prototypes of RISs as well. In [4], an RIS architecture is presented which can achieve amplitude-and-phase-varying modulation to facilitate MIMO transmission. The presented system is evaluated over-the-air (OTA) in real time while taking into account the hardware constraints and their impact on the system. In [5], the authors developed a new type of low-cost and high-gain RIS that comprises 256 two-bit elements. The RIS can provide 21.7 dBi antenna gain at 2.3 GHz and 19.1 dBi antenna gain at 28.5 GHz. In a recent study [6], OTA test results were presented for RIS-assisted communication systems operating in indoor and outdoor scenarios. The RIS comprises 1,100 tunable elements and can provide a 26 dB power gain in indoor scenarios and a 27 dB power gain in short-range outdoor tests. The RIS also provides a gain of 14 dB with a power consumption of only 1 W when the distance between the transmitter and the receiver is 500 meters. The first experiment for backscatter systems has also been recently conducted [7]. The proposed RIS is able to realize reflected beams (belonging to a codebook) that improve the tag-to-reader bit error rate.

As in-depth research activities on RISs are progressing, 17 white papers out mention RISs as a candidate technology for future networks, and 11 of them include a comprehensive analysis dedicated to RISs.

Funding agencies worldwide are providing more investment to the RIS technology by supporting fundamental and applied research works. Projects that are directly related to studying RISs for radio transmission are summarized in Table. I. Starting from 2012, RIS-dedicated research projects significantly contribute to the theoretical study and engineering development of the concept of RISs. Some projects are already completed with notable outcomes, like research papers and prototypes, while some are still ongoing and keep generating new ideas and concepts.

Within the IEEE Communications Society (ComSoc), two special interest groups (SIG) and one emerging technology initiative (ETI) have been established to foster collaborations between researchers in the particular field of study of RISs. Members of the SIGs and ETI are from different fields of study, such as wireless communications, signal processing, and metamaterials. In addition, these initiatives have attracted industrial players such as telecommunication operators and network infrastructure vendors, which provide insights from the industrial point of view and trigger discussions on the cost and potential applications of RISs.

Although we have witnessed extensive research efforts and activities dedicated to RISs globally, the standardization work has only taken off at a regional-level. In China, RIS-related standardization work has been kicked off in two different standard organizations, namely, the China Communications Standards Association (CCSA) and the FuTURE Mobile Communication Forum. CCSA is the official standard organization established by enterprises and institutes in China to carry out standardization activities in the field of Information and Communications Technology (ICT). During the 55th meeting of the technical committee 5 - working group 6 (TC5 WG6), the proposal to establish a study item (SI) on RIS was approved [8]. The SI started immediately after approval and will end in June 2022 with a technical report to be published. This SI aims to investigate numerous aspects which would help to realize RIS-assisted wireless communication systems, including channel modeling, channel estimation and feedback, beamforming with RISs, RISs and AI, and networking protocol for RIS-assisted networks. In the FuTURE forum, a working group to study how to integrate RISs into next generation wireless networks was established in December 2020 [9]. This working group will study the potential use cases and key technologies to support RISs to standardization and commercialization, and will publish a white paper to summarize the technical trends for further promoting RIS-assisted wireless systems. Within the European Telecommunications Standards Institute (ETSI), a new industry specification group (ISG) on RISs has been proposed and approved in June 2021. The proposal is led by founding members comprising chipset vendors, network infrastructure vendors, operators, research institutes and universities, and it aims to start working from September 2021. The planned output of this ISG includes technical reports, white papers and proof-of-concepts.

TABLE I
SUMMARY OF FUNDED RESEARCH PROJECTS ON RIS WITH EMPHASIZE ON WIRELESS TECHNOLOGIES

| Title | Main objectives and outcomes | Start month | End month | Funding agencies | Overall budget |
|---|---|---|---|---|---|
| Manipulating Terahertz Waves Using Three-Dimensional Metamaterials | Manipulation of the polarization of THz waves using passive or active metamaterial based devices. | September 2012 | August 2016 | NSF | $ 267,000 |
| VisorSurf – A Hardware Platform for Software-driven Functional Metasurfaces | Joint hardware and software designing for RIS with two experimental prototypes proposed. | January 2017 | December 2020 | Horizon 2020 | € 5,748,000 |
| Assessing the Feasibility of Programming the Ambient Wireless Environment | Exploring ways of changing the perceived channel along the wireless link to create more favorable conditions for communication. | May 2018 | April 2020 | NSF | $ 191,278 |
| Liquid Metal Tuned Flexible Metasurfaces | Deployable, transportable and conformable meta-surfaces that can be tuned on-demand and in real-time for radio signals. | September 2019 | August 2022 | NSF | $ 290,999 |
| Artificial Intelligence Aided D-band Network for 5G Long Term Evolution (ARIADNE) | Merging a novel high-frequency advanced radio architecture with an artificial intelligence network processing and management approach into a new type of intelligent communications system beyond 5G. | November 2019 | October 2022 | Horizon 2020 | € 5,968,393.75 |
| Scaling WLANs to TB/sec: THz Spectrum, Architectures, and Control | A first-of-its-kind pixelated metasurface waveguide to dynamically steer a THz beam via electrical switching of the meta-elements. | July 2020 | June 2025 | NSF | $ 171,453 |
| Enabling Seamless Coexistence between Passive and Active Networks using Reconfigurable Reflecting Surfaces | Developing a novel framework that enables seamless co-existence among multiple passive and active wireless systems, by exploiting the RIS concept to suppress interference at the wireless receivers. | October 2020 | September 2023 | NSF | $ 320,000 |
| Future Wireless Communications Empowered by Reconfigurable Intelligent metamaterials (META WIRELESS) | Enabling the manipulation of wireless propagation environments by reconfigurable intelligent surfaces. | December 2020 | November 2024 | Horizon 2020 | € 3,995,128.44 |
| Reconfigurable Intelligent Sustainable Environments for 6G Wireless Networks (RISE-6G) | Several of studies, engineering and standardization practices which can bring the technically advanced vision on RIS into industrial exploitation. | January 2021 | December 2023 | Horizon 2020 | € 6,499,613.75 |
| Harnessing multipath propagation in wireless networks: A meta-surface transformation of wireless networks into smart reconfigurable radio environments (Pathfinder) | Setting the theoretical and algorithmic bases of RIS-empowered wireless 2.0 networks that will lead to further transformations of wireless networks. | May 2021 | April 2023 | Horizon 2020 | € 184,707.84 |
| Surface waves in smart radio frequency environments (Surfer) | Pioneer the theoretic foundation and experimental validation of surface wave communications for indoor communications. | February 2022 | January 2024 | Horizon 2020 | € 184,707.84 |

The regional efforts to standardize RISs have now echos on global platforms. During the ITU-R WP 5D meeting in October 2020, countries submitted their candidate proposals for next generation wireless technologies, which will be considered for inclusion into the IMT Future Technology Trends report [10]. The WP 5D has the prime responsibility within ITU-R for issues related to the terrestrial component of IMT, including technical, operational and spectrum-related issues to meet the objectives of future IMT systems. During the meeting, companies from China submitted a proposal on the preliminary draft new report that included RISs as a key technology for future wireless networks. It is anticipated that more countries will mention, in future meetings, RISs in their proposals.

## III. Key Use Cases of RIS

Thanks to the ability of adjusting the reflection amplitudes and phase shifts of the impinging signals, RISs are able to reconfigure the end-to-end wireless channels, which brings new degrees of freedom to improve the wireless network performance. As such, RISs have a great potential for a wide range of applications, as shown in Fig. 1, including unmanned aerial vehicles (UAV) communications, mmWave coverage extension, wireless information and power transfer, physical layer security. For example, RISs can be either deployed on the ground to assist UAV communications with ground nodes or they can be attached to UAVs to assist terrestrial communications by leveraging smart passive reflections from the sky. Furthermore, by exploiting RISs, extra signal paths can be reflected towards desired directions to improve the channel condition rank, which is very crucial for high-speed MIMO transmissions with multiple data streams and to unleash the full potential of traditional active antennas at the BSs/APs. Besides, to improve the efficiency of simultaneous wireless information and power transfer (SWIPT), the large aperture of RISs can be leveraged to compensate for the significant power loss over long distances and to serve nearby massive Internet-of-things (IoT) devices. In traditional wireless networks with severe interference, the reflections of RISs properly deployed at the cell-edge can be used to reduce or even suppress the co-channel interference and inter-cell interference.

RISs are also expected to support the three pillar use cases of current 5G networks, namely, enhanced mobile broadband (eMBB), ultra-reliable and low-latency communications (URLLC), and massive machine-type communications (mMTC). For example, RISs can be deployed to improve the coverage and to establish high-capacity hotspots, which is crucial for eMBB and mMTC applications in stadiums, smart factories, shopping centers, and airports. In particular, in mMTC scenarios where the devices may switch from the active to the inactive mode frequently, the device activity detection efficiency and accuracy can be effectively improved by exploiting the additional paths provided by the RISs. Furthermore, RISs can be applied to effectively compensate for the Doppler and delay spread effects, which is important in the context of URLLC applications, such as smart transportation systems [11]. Specifically, exploiting RISs to turn typically random wireless channels into more deterministic ones can effectively reduce the number of retransmissions and can minimize the latency. Therefore, RISs are a revolutionary technology to upgrade the current infrastructure, which is expected to boost a wide range of industries and eventually helps achieve the smart home and smart city of the future.

## IV. Challenges from the Industrial Viewpoint

While RISs have a tremendous potential for improving the system capacity and the coverage area of next generation cellular networks, there are a number of technical challenges to be solved before they can be widely deployed in practical networks.

### A. Electromagnetic-Consistent Channel Models

RISs aim to improve the radio environment to make the propagation channels more favorable for communications. Thus, channel modeling becomes fundamental for studying RISs and for their performance evaluation. Channel modeling for RISs is more complicated than the traditional transmitter-to-receiver set-up in the sense that the electromagnetic characteristics of each RIS element needs to be precisely modeled. The challenge is also exacerbated by the large size of typical RISs, leading to peculiar near-field effects that would not normally be prominent in traditional wireless systems.

Some initial efforts have been carried out in free-space environments [12], where the classical Friis formula is assumed for each link connecting the transmitter to the RIS, and the RIS to the receiver. Each RIS element is modeled by a radiation pattern and a gain factor, while more complicated inter-element coupling effects are ignored. Such simplified model may be adequate for LOS scenarios, for instance, between a base station and an RIS, where they can be mounted well above the ground and their physical locations can be optimized. However, the link between an RIS and a terminal would often see significant non-line-of-sight (NLOS) components due to channel reflections, diffractions and/or building penetration losses. Therefore, more accurate propagation models should be considered, especially for RIS-terminal links. The modeling of RIS elements needs to be more sophisticated in order to capture the electromagnetic properties of each element and the coupling between adjacent elements of the RIS. Vast amounts of channel measurements are needed to appropriately characterize RISs in sub-6GHz and millimeter wave bands.

### B. Comparison with Relays

Generally speaking, relays can also shape a wireless propagation environment, although requiring active components, e.g., power amplifiers. As the predecessor of RISs, relays have, however, not yet gained momentum in the wireless industry. There exist mainly two types of relays: repeaters and layer 3 (L3) relays. A repeater simply amplifies the received RF signal (including noise) and forwards it to the other end. An L3 relay is equipped with physical layer and MAC layer processors, and decodes the received RF signal before forwarding a regenerated signal (often in a different format) to the end receiver, based on the decision of its own scheduler. An L3 relay is essentially a small-scale base station with a wireless backhaul. Unfortunately, L3 relays may have a quite limited use because of their rather complex physical layer protocols and expensive equipment. Repeaters are typically much cheaper than L3 relays and are also standards-transparent. The hardware cost can be further reduced in half-duplex repeaters, e.g., the input link and the output link are time multiplexed. However, repeaters are only used in certain areas affected by coverage holes.

The comparison between RISs and repeaters is more relevant, since the signals in both implementations are forwarded without being digitally processed. The tradeoff between RISs and repeaters id determined by (1) the cost of the active components of repeaters and the small unit gain of the nearly-passive RIS elements, and (2) the reduced digital control of repeaters and the control overhead of RISs for achieving the necessary beamforming gain. A preliminary performance comparison between RISs and repeaters is carried out in [13], where the achievable rates vs. the distance between an RIS and a half-duplex repeater are compared. An alternating optimization algorithm is proposed to maximize the achievable rates, by jointly optimizing the transmit beamforming and the reflection coefficients of the RIS. The ITU Urban Micro pathloss model and a Rayleigh small-scale fading channel are considered. The carrier frequency is 3 GHz and the bandwidth is 100 MHz. Both the base station and the repeater have 43 dBm transmit power and are equipped with 4 transmit antennas. The RIS has 200 elements. The channel state information (CSI) of each link is assumed to be known. From Fig. 2, it is seen that, in the considered single-cell and single-user setup, the achievable rates of an RIS are comparable to those of a half-duplex repeater.

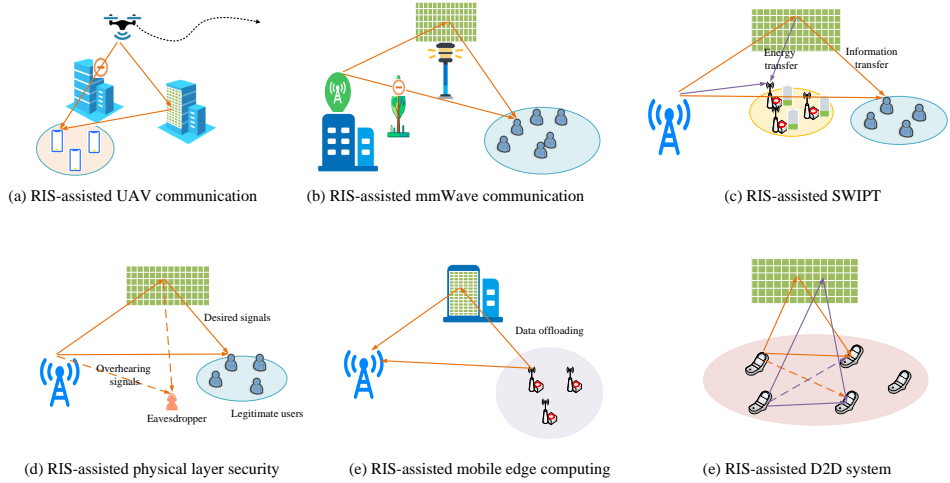

(a) RIS-assisted UAV communication  (b) RIS-assisted mmWave communication  (c) RIS-assisted SWIPT

(d) RIS-assisted physical layer security  (e) RIS-assisted mobile edge computing  (e) RIS-assisted D2D system

Fig. 1. Illustration of representative RIS-aided use cases in future wireless networks.

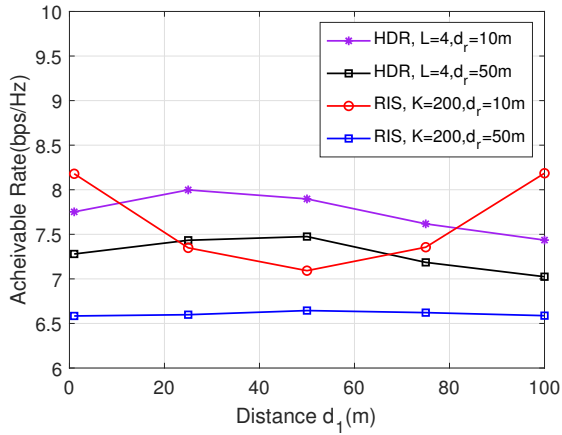

Fig. 2. Achievable rates vs. distance for an RIS and a half-duplex repeater.

### C. Practical Issues at the Physical Layer

As a nearly-passive device, an RIS is not supposed to generate and transmit its own signals, although it may be able to decode the control signals. This poses some challenges for channel estimations. For example, only the cascaded channel, i.e., the product of the transmitter-to-RIS and RIS-to-receiver link that includes the RIS response, can be measured. If efficient channel estimation algorithms and protocols are not used, the channel estimation phase may require a significant overhead. Also, different from active RF devices, the electric characteristics of the RIS elements may be far from ideal and the control range of the phases and their quantization levels may be limited.

For dynamic scheduling, the presence of an RIS needs be known to the base stations for properly controlling them. An RIS may or may not be transparent to terminals. In the latter case, more control signaling is needed for the terminals to choose an RIS. Equipped with a large number of elements, an RIS can be seen as a multi-antenna device and it needs to work jointly with multi-antenna base stations and mobile terminals. This calls for more complicated algorithms for resource schedulers.

To enable the flexible deployment of RISs, the inherent limitations of wired lines and power lines for controlling them cannot be under-estimated. In addition to the design of efficient algorithms for channel estimation and signal processing, the control signaling is a key design aspect that needs to be realized at low power consumption and low complexity.

### D. Control of the Frequency Bands

RISs do not usually have digital or RF chains to process the incident signals, and, thus, they cannot discriminate signals at a fine granularity as a function of the frequency band. This results in at least two issues. First, if the allocated bands of two operators are adjacent, the passive beamforming of an RIS that belongs to an operator will inadvertently disrupt the operation of the other operator's network and will cause additional interference. Second, the granularity of the frequency selection of an RIS may not be as fine as several hundred kHz as for an active RF filter. This may impose certain limitations on the performance potential of RISs in terms of frequency-selective scheduling.

*E. Manufacturing of Metasurface-based RISs*

An RIS needs to be large enough to compensate for the link budget deficit that originates from its nearly-passive implementation, e.g., the absence of power amplifiers. In general, an RIS needs to be equipped with hundreds of elements to offer a competitive gain. To make the implementation and deployment of RISs affordable, each element needs to be low-cost.

The reflecting elements of an RIS need to have consistent electric properties, such as the reflection index, over a wide range of angles of incidence and reflection, so that a single RIS can serve multiple terminals in different locations of a cell.

RISs may be deployed in outdoors, e.g., on a building facade, on a billboard, etc. Outdoor environments are prone to harsh operating conditions, such as the sunlight, temperature fluctuations, wind, rain, ice/snow, pollution. The materials that constitute the RIS elements and their associated control circuitry need to be resilient to these adverse weather conditions for months or years, during which their electromagnetic properties should not significantly deteriorate.

## V. STANDARDIZATION OF RISS: FORECASTED ROADMAP

So far, we have mainly discussed the integration of RISs in cellular networks. To clearly understand the way forward towards the future commercial deployment of RIS-assisted wireless networks, a critical analysis of the associated standardization process is necessary. The typical path towards standardizing an emerging technology, usually follows a well-defined process.

The emerging technology is usually first thoroughly studied by academic researchers for a considerable period of time, during which fundamental solutions to technical problems are given and initial use cases are identified. In the meantime, the emerging technology is investigated by the industry in one or more major use cases and practical scenarios. As both academic and industrial researchers proceed with their research activities, the first prototypes are developed and tested, which leads to a fast paced development of the technology. It is at this stage that early standardization activities are initiated, in order to drive the progress of commercialization and industrialization. The standardization process usually begins with study items (SIs) in one or multiple regional standards developing organizations (SDO) or industrial fora, and proceeds in international SDOs, like the 3rd Generation Partnership Project (3GPP). The next step is the instantiation of a work item (WI), which leads to technical specifications that are comprehensively defined and then released. This last step is perceived as the sign that the technology is officially part of the global standards.

With substantial investments from research institutes and businesses, a fast progress in terms of implementing and testing advanced and realistic prototypes is expected. As depicted in Fig. 3, corporations in the wireless industry have already started the realization of prototypes since the year of 2020. The tests executed provide researchers and engineers with firsthand data on the performance offered by RISs in cellular networks.

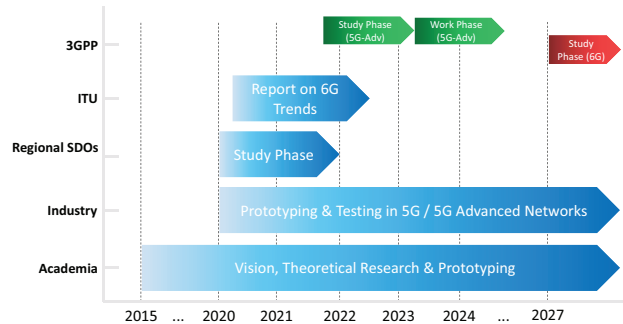

Fig. 3. Forecasted timeline for the standardization of RISs.

In regional SDOs, the study of RISs is under its way already. Usually, SIs and WIs last for two years, and their outcomes are usually technical reports or white papers. Within the ITU, one critical milestone is the official release of a report for 6G trends which is usually perceived as a guidebook for next generation wireless networks. The discussion on this latter report started in February 2020, and the report is expected to be published in June 2022. According to the draft report already available in [14], RISs are described as a critical component for the physical layer of 6G networks. Since the ITU focuses on regulatory, spectrum and business aspects, it is unlikely for RIS to be established as a focus group or a WI there. Further actions in the ITU will become clearer after 2023.

As far the 3GPP is concerned, the current focus is on completing Release 17 for 5G systems, with possible enhancements in Release 18, but there is no formal plan for 6G systems. According to past experience, each generation of wireless communication standards lasts for approximately a decade. The first 3GPP working group meeting for 5G systems started in April 2016, indicating that it is reasonable to expect that possible discussions on 6G systems will begin after 2026.

Generally speaking, there are two possible ways to standardize RISs in 3GPP. One is to first initiate an SI on scenarios and channel models in Release 18 and then to initiate a WI in Release 19. This would allow for the deployment of RISs as a component of 5G advanced networks. The advantage of this approach is that RISs can be deployed in wireless networks in a massive scale soon, which will help maturing this technology. In this case, RISs will be part of 5G standards and, therefore, they will be automatically included in 6G standards. Another possible path may be to standardize RISs as part of 6G standards together with other new features for 6G. This implies that the deployment of RISs may be delayed to the next generation networks, which is usually subject to uncertainties. Since the RIS technology can be seen as generic and band-agnostic, standardizing it as early as possible would enable opportunities for RISs to be combined with other technologies, thus contributing to greener, safer and more reliable wireless networks. On the other hand, there are still some challenging technical problems to be tackled, which require considerable amount of time. In this sense, the exact date to start officially considering RISs in 3GPP largely depends on the maturity of this technology. Since there was a first proposal submitted to 3GPP during the March 2021 meeting [15], the earliest possible time to start an SI on RISs is in the late 2021.

At the time of writing, the first concrete milestone towards the possible integration of the RIS technology into future communication standards is the official approval in June 2021 of an ETSI industry specification group (ISG) on RISs. The ISG-RIS will kick off the activities in September 2021 and its mission is to provide an opportunity to coordinate pre-standards research efforts on the RIS technology across various collaborative projects towards paving the way for future standardization of the technology.

## VI. Conclusions

RISs provide opportunities to manipulate the propagation channels and to enable smart radio environments at a reduced power consumption and cost. The associated research challenges include channel modeling, channel estimation and feedback, as well as the real time smart control of RISs. These challenges are being addressed by collaborative research projects funded by national and international agencies, as well as the industry. Looking ahead, along with the theoretical and technological enhancements which will improve the performance of RISs while reducing the associated implementation and deployment cost, standardization efforts are needed to determine appropriate channel models, waveforms, control signalling, and performance requirements considering the currently available spectrum, technology, and test solutions. As RISs start to be integrated into wireless networks, new use cases will be identified and further enhancements to this technology will be proposed towards the realization of truly controllable, programmable, and smart radio environments.

**Ruiqi Liu** (richie.leo@zte.com.cn) is currently a senior researcher in ZTE Corporation. He serves as the co-rapporteur of the work item on NR RRM enhancement in 3GPP, the Editor of ITU Journal of Future and Evolving Technologies and the Standards Liaison Officer for IEEE ComSoc Signal Processing and Computing for Communications Technical Committee.

**Qingqing Wu** (qingqingwu@um.edu.mo) is currently an assistant professor with the State key laboratory of Internet of Things for Smart City, University of Macau.

**Marco Di Renzo** (marco.di-renzo@universite-paris-saclay.fr) is a CNRS Research Director with the French National Center of Scientific Research in CentraleSupelec, Paris-Saclay University, France. He currently serves as the Editor-in-Chief of IEEE Communications Letters and is an IEEE Fellow.

**Yifei Yuan** (yuanyifei@chinamobile.com) was with Alcatel-Lucent from 2000 to 2008. From 2008 to 2020, he was with ZTE Corporation as a technical director and chief engineer, responsible for standards & research of LTE-Advanced and 5G technologies. He joined the China Mobile Research Institute in 2020 as a Chief Expert, responsible for 6G. He has extensive publications, including seven books on LTE-Advanced and 5G. He has over 60 granted patents.